\documentstyle[12pt]{article}

\begin{document}

\title{DARK BARYONS IN GALACTIC HALOS} 

\author{MARCO RONCADELLI \\ 
INFN, Pavia, Italy (roncadelli@pv.infn.it) }

\date{}

\maketitle




\begin{abstract}
Primordial nucleosynthesis as well as anisotropies in the 
cosmic microwave background radiation imply that the total amount of baryons 
in the Universe largely exceeds the visible contribution, thereby making a 
strong case for {\it baryonic dark matter}. Moreover, certain recent 
developments lead to a consistent picture of the dark baryon budget in the 
present-day Universe. Accordingly, {\it dark 
baryons are mostly locked up in galactic halos} -- which are anyway dominated 
by nonbaryonic dark matter -- and {\it a sizable fraction of them consists of 
gas clouds}. While {\it a priori} various forms of baryonic dark matter in 
galaxies can be conceived, observational constraints rule out most of the 
possibilities, leaving 
{\it brown dwarfs} and {\it cold gas clouds mostly made of $H_2$} as the only 
viable candidates (besides 
supermassive black holes). So, it looks natural to suppose that baryonic dark 
matter in galaxies is accounted for by {\it dark clusters} made of brown 
dwarfs and 
cold $H_2$ clouds. A few years ago, it was shown that indeed these dark 
clusters are predicted to populate the outer halos of normal spiral galaxies 
by the 
Fall-Rees theory for the formation of globular clusters, which was based on 
the standard cold dark matter paradigm described in Blumenthal et al. 1984 
Nature 311, 517. We review the dark cluster formation mechanism, and argue 
that its {\it qualitative features} are expected to remain true even in the 
contemporary 
picture of galaxy formation. We also discuss various ramifications of 
the dark cluster scenario in question, paying particular attention to its 
observational implications. One of them -- the 
diffuse gamma-ray emission from the Milky Way halo -- appears to have been 
confirmed by the discovery of Dixon et al. 1998 New Astronomy 3, 539. Whether 
this is actually fact or fiction only the future satellite missions {\it 
AGILE} and {\it GLAST} will tell.
\end{abstract}

\section{Introduction and motivation}

All available observational evidence -- including flat rotation curves of 
spiral galaxies, X-ray emission from elliptical galaxies and clusters of 
galaxies, gravitational lensing by galaxy clusters, velocity fields in deep 
galaxy surveys, light curves of high-redshift Type-Ia supernovae and cosmic 
microwave background radiation (CMBR) anisotropies -- invariably lead to the 
conclusion that the Universe is dominated by dark matter. Denoting (as 
usual) by 
$\Omega$ the actual density of the present-day Universe in units of the 
critical 
density, the situation can be summarized as follows. 

Measurements of anisotropies in the CMBR yield both the {\it total} value 
$\Omega \simeq 1$ from the location of the first acoustic peak~\footnote{
Observe that this result agrees with the natural 
implication of cosmic inflationary scenarios (Liddle \& Lyth 2000).} (De 
Bernardis et al. 2002; Balbi et 
al. 2000; Pryke et al. 2002) and the {\it baryonic} contribution from the 
height of the second peak~\footnote{Behind this second conclusion there is the 
assumption that primordial fluctuations are adiabatic, which is just what 
cosmic inflation entails. The value $H_0 \simeq 70 \, km \, s^{- 1} \, 
Mpc^{- 1}$ for the Hubble constant is adopted throughout.} ${\Omega}_b 
\simeq 0.04$ (Sievers et al. 2002; Stompor et al. 2001). Quite remarkably, 
the latter value is in good agreement with two {\it independent} estimates. 
One arises by combining measurements of the primeval abundance of light 
elements (especially deuterium) with theoretical predictions from primordial 
nucleosynthesis (Burles et al. 2001; O'Meara et al. 2001). The other follows 
from the observed features of {\it high-redshift} Lyman-$\alpha$ forest 
absorption lines of neutral hydrogen (Rauch et al. 1998) observed in the 
spectra of background quasars~\footnote{Lyman-$\alpha$ forest lines will be 
addressed in more detail in Sect. 3.}. However, this is not the end of the 
story, since estimates of the total {\it matter} distribution -- based on the 
spectral features of the CMBR, on deep galaxy surveys and on various 
studies of clusters of galaxies -- yield ${\Omega}_m \simeq 0.3$ 
(Turner 2002). Evidently, matter alone -- as clustered on various cosmic 
scales -- fails to account for the total energy budget of the Universe, 
leaving a gap of ${\Omega} - {\Omega}_m \simeq 0.7$. Actually, the situation 
is even more puzzling, since observations of high-redshift Type-Ia supernovae 
entail that the present cosmic expansion is accelerated. Nevertheless, this 
behaviour is explained by a {\it nontrivial vacuum} (unclustered energy) with 
negative pressure and ${\Omega}_{\Lambda} \simeq 0.7$~\footnote{This vacuum is 
described by a cosmological constant ${\Lambda}$.}~(Riess et al. 1998; 
Perlmutter et al. 1999). Notice that such a vacuum provides {\it just} the 
missing contribution ${\Omega} - {\Omega}_m \simeq 0.7$ mentioned above, and 
so the overall picture is indeed remarkably {\it consistent}. 

The foregoing discussion implies ${\Omega}_b/{\Omega}_m \simeq 0.13$, 
thereby showing that most of the dark matter is nonbaryonic. This is in 
agreement with contemporary theories of galaxy formation, which demand that 
the observed dark galactic halos should be dominated by nonbaryonic matter.

Still -- as we will see below --  {\it half} of the total amount of baryons 
in the present-day Universe remains observationally {\it unaccounted}. Hence, 
there is also 
a {\it baryonic dark matter problem}, which is the main topic addressed in 
this paper.

An inventory of the baryonic content of the various cosmic structures has 
been attempted by Persic and Salucci (1992), Gnedin and Ostriker (1992), 
Bristow and Phillipps (1994), and Fukugita, Hogan and Peebles (1998) by 
combining available observational data with the whole body of theoretical 
knowledge. According to the latter authors, the baryon budget in the 
present-day Universe yields ${\Omega}'_b \simeq 0.02$ and is dominated by 
hot gas in groups and clusters of galaxies, while the equivalent cosmological 
contribution from luminous baryons~\footnote{Luminous baryons include stars, 
atomic and molecular hydrogen whose existence is inferred from detection of the 
emitted electromagnetic radiation at any wavelenght.} in galaxies is as small 
as ${\Omega}'_{bg} \simeq 0.004$. 

Yet, we know that the observed CMBR anisotropies, primordial nucleosynthesis 
and high-redshift Lyman-$\alpha$ forest absorption require ${\Omega}_b \simeq 
0.04$. {\it Where are the dark 
baryons}?

An answer to this question is provided by hydrodynamic simulations of 
currently favoured structure formation models~(Cen \& Ostriker 1999; Dav\`e 
et al. 1999). They indicate 
that -- in the present-day Universe -- baryons are divided {\it almost 
equally} among three phases: (i) {\it warm gas} $(T = 10^5 - 10^7 \, K)$, 
forming transient filamentary structures in intergalactic space; (ii) 
{\it cool gas} $(T < 10^5 \, K)$, smoothly diffuse in intergalactic space; 
and (iii) {\it cold gas} and {\it stars} associated with galaxies~(Dav\`e et 
al. 2001). Although the uncertainties in the simulations prevent a precise 
estimate of the fractional abundance of baryons in the three phases, the 
equivalent cosmological density in each one is roughly $0.013$. This means 
that the equivalent cosmological contribution from galactic baryons should be 
${\Omega}''_{bg} \simeq 0.013$. Because we have seen that luminous baryons in 
galaxies only give ${\Omega}'_{bg} \simeq 0.004$, we are led to the conclusion 
that in the low-redshift Universe {\it most of the galactic baryons are dark}.

Actually, a strong case for a {\it somewhat larger} amount of baryonic dark 
matter in galaxies has recently been made by Kochanek (2001). He has 
investigated the mass function of the dark matter halos -- as predicted by the 
current cold dark matter paradigm -- by using certain dynamical probes (Kochanek 
\& White 2001). More specifically, Kochanek has shown that the 
distribution of gravitational lens separations and the circular velocity 
function predicted by semi-analytic models of hierarchical clustering are both 
in agreement with observations {\it only} provided that ${\Omega}_{bg} 
\simeq 0.02$~\footnote{We remark that no serious conflict exists between this 
conclusion and the outcome of hydrodynamic simulations, owing to their 
intrinsic uncertainties.}. Basically the same conclusion has been reached by 
Klypin, Zhao and Somerville (2001) via N-body simulation modelling of the 
Milky Way and Andromeda galaxy. These authors found that the observed 
properties are reproduced {\it only} if $50 \, \% - 75 \, \%$ of the galactic 
baryons are {\it nonluminous}. Consequently, {\it galactic halos 
are the main repository of dark baryons} at low redshift~\footnote{Of course, 
this does not undermine the fact that galactic halos are dominated by 
nonbaryonic dark matter.}~\footnote{Strictly speaking, there is a logical gap 
here, which should be discussed. Contrary to hydrodynamical simulations, 
N-body simulations and semi-analytical models of galaxy formation neglect 
feedback processes, which give rise to  galactic outflows. Therefore, the 
above results of Kochanek and Klypin, Zhao and Somerville do not prevent 
ejection of a large amount of baryons during the galaxy evolution. 
Nevertheless, there are good reasons to believe that most of the baryons 
should still lie inside galactic halos {\it today}. For, galactic winds 
powered by supernova explosions appear unable to drive a huge amount of gas 
out of galactic halos (Mac Low \& Ferrara 1999). In addition, a large amount 
of ejected baryons from spiral galaxies would produce an equivalent 
cosmological density of the intergalactic medium larger than the one predicted 
by the above-mentioned hydrodynamical simulations.}.

As a matter of fact, this circumstance is in remarkable agreement with the 
claim of Lanzetta and collaborators ~(Chen et al. 1998, 2001) that 
{\it low-redshift} Lyman-${\alpha}$ forest absorption lines of neutral 
hydrogen are produced by {\it baryonic gas clouds} located inside the dark 
galactic halos. Notice that these gaseous baryons are {\it missing} in the 
above-mentioned inventories.

Apparently, a consistent picture of the dark baryon budget in the 
low-redshift Universe emerges from the previous considerations. {\it Dark 
baryons are mostly locked up in galactic halos}, and {\it a sizable fraction 
of them consists of gas clouds}.

Several kinds of astronomical objects -- like {\it brown dwarfs}~\footnote{We 
recall that brown dwarfs are stellar-like objects in which ordinary nuclear 
reactions do {\it not} occur in the core, because their central temperature 
never exceeds the hydrogen-burning threshold (for a review, see Kulkarni 1997; 
Basri 2000; Chabrier \& Baraffe 2000). In the case of solar metallicity, the 
brown dwarf mass range is 
$0.01 - 0.08 \, M_{\odot}$, but for primordial metallicity the upper limit 
becomes $M < 0.11 \, M_{\odot}$~(D'Antona 1987). For a long time, brown dwarfs 
were regarded merely as a theoretical possibility, until they were first 
discovered in $1995$ (Rebolo et al. 1995; Nakajima et al. 1995). Further 
observations have shown 
that brown dwarfs form like ordinary stars. Similarly to what happens in very 
low-mass stars (Drake et al. 1996), it has been argued that also in brown 
dwarfs a turbolent dynamo process should heat the corona, which consequently 
emits soft X-rays (Kashyap et al. 1994; Neuh\"auser et al. 1999). This point 
will be important in some later considerations.}, {\it red dwarfs, white 
dwarfs, neutron stars, black holes} and {\it gas clouds} -- may account for 
baryonic dark matter in galaxies 
(Carr 1994), but observational constrains rule out most of them. Before 
a star becomes a {\it white dwarf}, a {\it neutron star} 
or a {\it black hole}, it goes through phases in which heavy elements are 
produced and ejected into interstellar space and its luminosity is greatly 
enhanced. Existing upper 
limits on the metallicity of the environment and on the background light then 
rule out these objects as candidates for baryonic dark matter~\footnote{The 
metallicity constraint is evaded by black holes with mass $> 200 \, 
M_{\odot}$, because they would ingurgitate their whole parent stars. However, 
only supermassive black holes -- with mass $> 10^5 \, M_{\odot}$ -- turn 
out to be viable dark matter candidates, for otherwise the parent stars would 
violate background light constraints.} (Carr 2000; Madau et al. 2000). A 
similar conclusion holds for {\it red dwarfs} as well, because they would 
give rise to infrared fluxes 
stronger than are detected (more about this, in Sect. 3). As far as gas 
clouds are concerned, only {\it cold gas mostly made of $H_2$} is a viable 
possibility, for 
otherwise an unseen excess of photons at some wavelenght ($21 \, cm$ for 
neutral hydrogen) would be produced~\footnote{As $H_2$ does not possess a 
permanent electric dipole moment, the first transition line to be detected in 
emission occurs by quadrupole radiation at $512 \, K$ above the ground state. 
Thus, cold $H_2$ clouds do not emit any radiation~(Combes \& Pfenniger 
1997). Notice that astronomers tend to infer the existence of $H_2$ by 
using $CO$ as a tracer. Needless to say, this practice makes sense {\it only} 
when the ${H_2}/{CO}$ conversion ratio is under control (like in the Milky 
Way disk), but this is {\it not} the case in galactic halos. The possibility 
that a large amount of $H_2$ may be present in the Universe was first 
suggested by Zwicky (1959).}. So -- with the exception of supermassive black 
holes --  only {\it brown dwarfs} and {\it cold $H_2$ clouds} turn out to be 
{\it realistic} forms of baryonic dark matter in galaxies~\footnote{Of 
course, a small fraction of red dwarfs, white dwarfs, neutron stars and black 
holes can be present in the galactic halos, but they cannot account for the 
bulk of baryonic dark matter.}~\footnote{A strong constraint on the brown 
dwarf fraction in galactic halos -- which would prevent them from playing any 
r\^ole for dark matter -- was derived by Graff and Freese (1996a) under the 
assumption that the initial mass function is spatially {\it homogeneous}. 
However, within the dark matter model to be discussed below the initial mass 
function {\it depends} on the galactocentric distance, and brown dwarfs can 
copiously form only at galactocentric distances larger than $10 - 20 \, kpc$. 
Hence, the conclusion of Graff and Freese (1996a) is {\it not} valid in the 
present context.}~\footnote{We anticipate (from Sect. 3) that {\it old} brown 
dwarfs are {\it not} ruled out by current infrared searches (especially if 
they are clustered).}.

Our aim is to discuss the possibility that the dark galactic baryons indeed 
consist of {\it dark clusters} -- made of {\it brown dwarfs} and {\it cold 
$H_2$ clouds} -- lurking in the halos of normal spiral galaxies.

More specifically, we will focus our attention on a model of this kind, which 
was developed from first principles by De Paolis, Ingrosso, Jetzer and 
Roncadelli (1995a,b, 1996, 1998a) within the standard cold dark matter 
picture of galaxy formation. We stress that an almost identical scenario 
has been independently proposed and investigated by Gerhard and Silk~(1996). 
Somewhat similar ideas have been put forward by Ashman and Carr~(1988), 
Ashman~(1990), and Fabian and Nulsen~(1994, 1997). Slightly different baryonic 
pictures have been suggested by Pfenniger, Combes and Martinet~(1994), 
Wassermann and Salpeter~(1994), Gibson and Schild~(1999a,b), Sciama~(2000), 
and Lawrence (2001).

Here, we offer an updated review of the above-mentioned dark matter model. 

The paper is organized as follows. In Sect. 2, we discuss the formation 
mechanism (Subsect. 2.1) and the main properties (Subsect. 2.2) of the 
dark clusters. Next, we address some of their observational 
implications in Sect. 3. Finally, we summarize our conclusions in Sect. 4. 
Suggestions for further work will be scattered throughout the paper.

\section{Dark clusters in galactic halos}

In the past, brown dwarfs have been repeatedly recognized as an important 
contributor to baryonic dark matter in galaxies (Carr 1994). Much like 
ordinary stars, brown dwarfs are expected to form in 
clusters, thereby giving rise to {\it dark clusters}~(Carr \& Lacey 
1987; Silk 1991; Rix \& Lake 1993; Kerins \& Carr 1994; Moore \& Silk 1995). 
However, 
insufficient attention was paid to the fact that -- even in this case -- 
the star-formation mechanism is highly inefficient, in that most of the 
original gas fails to get transformed into stars, thereby remaining trapped 
inside the dark clusters. As we shall see, this gas plays a crucial r\^ole, 
since it can both change the properties of the brown dwarfs and make the dark 
clusters directly visible in the gamma-ray band and indirectly observable 
along the line of sight to a distant quasar.

\subsection{Formation of the dark clusters}

Starting point of the considered dark matter model was the Fall-Rees (1985) 
theory for 
the formation of globular clusters in galactic halos. Indeed, it was shown 
(De Paolis et al. 1995a,b, 1996, 1998a) that the 
Fall-Rees theory {\it automatically} predicts -- without any further physical 
assumption -- that {\it dark clusters} made of brown dwarfs and cold $H_2$ 
clouds should form in the {\it outer halos} of normal spiral galaxies, namely 
at galactocentric distances larger than $10 - 20 \, kpc$. 

The Fall-Rees theory was formutated within the standard cold dark matter 
picture of galaxy formation described in Blumenthal et al. (1984). 
However, in the last few years numerical simulations of structure formation 
have led to a considerable improvement in the understanding of hierarchical 
clustering, with mergers playing an increasing r\^ole. As a result, some 
features of the above-mentioned cold dark matter paradigm turned out to be 
incorrect and had to be revised. Therefore, it is not clear whether the 
present dark matter model remains true. Although this 
is at the moment an open question, we will argue -- on the basis of the 
following discussion -- that the {\it qualitative features} of the dark matter 
scenario under consideration are likely to remain unchanged even in contemporary 
models of galaxy formation.

Consider a protogalactic cloud with mass $M \sim 10^{12} \, M_{\odot}$, made 
of both nonbaryonic and baryonic matter~\footnote{We suppose that the 
baryonic matter has primordial composition.}. We imagine that its inner part 
eventually produces -- upon gravitational collapse -- the luminous component 
of a normal spiral galaxy, whereas its outer region gives rise to the 
corresponding dark halo. 

The Fall-Rees theory rests on the assumption that the outer part of the 
protogalactic cloud is modelled by a {\it singular isothermal sphere} (SIS), 
so that the observed flat rotation 
curve is recovered. Accordingly, the {\it overall} density profile 
is $\rho_o(r) = V^2_{max}/4 \pi G r^2$, where $V_{max}$ is the value at which 
the rotation velocity becomes constant at a large galactocentric distance 
$r$. Typically one finds $V_{max} \sim 200 \, km \, s^{-1}$, and we will 
choose $V_{max} = 220 \, km \, s^{-1}$, which is the value appropriate to the 
Milky Way. Hence, we get
\begin{equation}
\rho_o(r) \simeq 0.9 \, \left( \frac{kpc}{r} \right)^2 \, M_{\odot} \, 
pc^{-3}~.   
\label{v1}
\end{equation}
Correspondingly, the free-fall time $t_{ff} \simeq 0.5 \, \alpha^{- 1/2} 
(G \rho_o)^{- 1/2}$ becomes 
\begin{equation}
t_{ff} \simeq 7.4 \cdot 10^6 \, \alpha^{- 1/2} \, \left( \frac{r}{kpc} \right) 
\, yr~,   \label{v2}
\end{equation}
where $\alpha$ is a constant of order unity which reflects the actual mass 
distribution (a pure SIS yields $\alpha = 1$). 
Moreover, the mean square velocity of any baryon in the considered 
region is $\overline{v^2} = 3 V^2_{max}/2 \simeq 7.3 \cdot 10^4 \, km^2 \, 
s^{-2}$. Therefore the temperature of the diffuse baryonic gas~\footnote
{As we will be explicitly concerned with the {\it baryonic} gas component 
{\it only}, 
the attribute baryonic will be implicitly understood. Also, the meaning of 
the attribute {\it diffuse} will become clear later on.} is $T_d 
\simeq 1.7 \cdot 10^6 \, K$. One can easily see that this temperature is 
quite close to the virial temperature, and so the diffuse gas is initially 
in virial equilibrium.

Since the diffuse gas tends to cool, in order to understand its further 
behaviour one has to compare the cooling time $t_{cool}$ with the free-fall 
time 
$t_{ff}$ as given by eq. (\ref{v2})~\footnote{We recall that a cloud collapses 
in free-fall and fragments when $t_{cool} < t_{ff}$, whereas the collapse 
proceeds quasi-statically for $t_{cool} > t_{ff}$.}. One can show that at the 
above temperature both Bremsstrahlung and ion-recombination processes 
contribute to the cooling rate of the diffuse gas, and the resulting cooling 
time is
\begin{equation}
t_{cool} \simeq 5.5 \cdot 10^{- 18} \, \left(\frac{g \, cm^{-3}}{\rho_d} 
\right) \, yr~,   \label{v3}
\end{equation}
where $\rho_d$ denotes the density of the diffuse gas. The comparison between 
eqs. (\ref{v2}) and (\ref{v3}) entails
\begin{equation}
\frac{t_{cool}}{t_{ff}} \simeq \frac{\rho_*}{\rho_d}~,   
\label{v4}
\end{equation}
where we have set
\begin{equation}
\rho_* \equiv 1.1 \cdot 10^{-2} \, \alpha^{1/2} \, \left( \frac{kpc}{r} \right) \, M_{\odot} \, pc^{-3}~.
\label{v5}
\end{equation}
On account of eq. (\ref{v4}), we would conclude that the diffuse gas undergoes 
quasi-static collapse as long as its density is sufficiently small -- namely 
for $\rho_d < \rho_*$ -- since then $t_{cool} > t_{ff}$. Still, as soon as 
$\rho_d > \rho_*$ the collapse enters the free-fall regime. Actually, in the 
latter case we would expect a monotonic increase of $\rho_d$, and -- thanks 
to eq. (\ref{v4}) -- a steady decrease of the $t_{cool}/t_{ff}$ ratio, 
which would signal efficient fragmentation and star formation within the 
{\it whole} protogalactic cloud. 

But in reality the situation is more complex, since the above picture would 
be correct only within a {\it perfectly homogeneous} medium. In fact, Fall 
and Rees 
have shown that the unavoidable presence of {\it density fluctuations} during 
the collapse can totally upset such an expectation. Basically, this is 
due to the fact that density fluctuations produce a {\it thermal instability} 
within the diffuse gas. To see this, consider eq. (\ref{v3}): an overdense 
region cools more rapidly than average, but then compression from the 
surrounding hotter gas leads to a further increase of the density. Thus, 
density fluctuations give rise to a {\it 
substructure} within the protogalaxy, made of denser and cooler gas 
{\it bubbles} embedded in the {\it diffuse} gas. Besides, Fall and Rees have 
shown that the bubble condensation implies the condition
\begin{equation}
\frac{t_{cool}}{t_{ff}} \simeq 1~   \label{v6}
\end{equation}
to hold true for the {\it diffuse} gas. This fact is extremely important, for 
three 
different reasons. First, the diffuse gas actually {\it stays} in virial 
equilibrium at $T_d \simeq 1.7 \cdot 10^6 \, K$. Second, there is {\it no} 
star formation within the 
{\it diffuse} gas. Third, eqs. (\ref{v4}), (\ref{v5}) and (\ref{v6}) 
imply that the density profile of the {\it diffuse} gas is
\begin{equation}
\rho_d(r) \simeq 1.1 \cdot 10^{-2} \, \alpha^{1/2} \, \left( \frac{kpc}{r} 
\right) \, M_{\odot} \, pc^{-3}~.
\label{v7}
\end{equation}

The bubbles -- which are pressure-confined by the diffuse gas -- are 
originally made of ionized hydrogen and helium (plus their respective 
electrons). Assuming that this plasma is in {\it thermal equilibrium}, the 
cooling brought about Bremsstrahlung and ion-recombination processes is 
operative only at temperatures larger than $\sim 10^4 \, K$, since at lower 
temperatures the gas becomes 
neutral and the corresponding cooling rate suddenly drops to zero. Hence, we 
expect that inside the bubbles {\it hydrostatic equilibrium} sets in when the 
temperature reaches $\sim 10^4 \, K$. In such a situation, the baryonic component 
of the protogalaxy is made of bubbles with $T_b \sim 10^4 \, K$ embedded in 
the diffuse gas at $T_d \sim 10^6 \, K$. 

Let us denote by $\rho_b(r)$ the (constant) density inside a bubble located 
at galactocentric distance $r$. At hydrostatic equilibrium, the pressure 
$p_b$ just inside the considered bubble coincides with the pressure $p_d$ 
just outside it (within the diffuse gas). This condition entails 
\begin{equation}
\rho_b (r) = \left( \frac{m_b}{m_d} \right) \left( \frac{T_d}{T_b} \right) \, 
\rho_d (r)~,   \label{v8}
\end{equation}
where $m_d \simeq 10^{- 24} \, g$ and $m_b \simeq 2 \cdot 10^{- 24} \, g$ are 
the mean particle mass of a ionized and neutral gas having primordial 
composition, respectively. Combining eqs. (\ref{v7}) and (\ref{v8}) together, 
we find
\begin{equation}
\rho_b (r) \simeq \, 3.6 \, \alpha^{1/2} \, \left( \frac{kpc}{r} \right) \, 
M_{\odot} \, pc^{-3}~.   \label{v9}
\end{equation}
Therefore the Jeans mass and the Jeans radius for the bubble in question are
\begin{equation}
M_J (r) = 
\left( \frac{3 k_B T_b}{\alpha G m_b} \right)^{3/2} \left( \frac{3}
{4 \pi \rho_b(r)} \right)^{1/2} 
\simeq 2.3 \cdot 10^6 \, \alpha ^{-7/4} \, 
\left( \frac{r}{kpc} \right)^{1/2} \, M_{\odot}
\label{v10}
\end{equation}
and
\begin{equation}
r_J (r) = 
\left( \frac{3 k_B T_b}{\alpha G m_b} \right)^{1/2} \left( \frac{3}
{4 \pi \rho_b(r)} \right)^{1/2} 
\simeq 54 \, \alpha ^{-3/4} \, \left( \frac{r}{kpc} 
\right)^{1/2} \, pc~,
\label{v11}
\end{equation}
respectively, where $k_B$ denotes the Boltzmann constant.

However, the above assumption -- namely thermal equilibrium of the plasma 
inside the bubbles -- turns out to be {\it violated}, since 
cooling occurs very rapidly~(Kang et al. 1990). The resulting {\it 
out-of-equilibrium} ion 
recombination entails a substantial ionization fraction at $T_b < 10^4 \, K$ 
(which would otherwise vanish at thermal equilibrium). This circumstance is 
in fact irrelevant for $T_b > 10^4 \, K$, but alters dramatically the above 
conclusions for $T_b < 10^4 \, K$. Indeed, the presence of a nonnegligible 
amount of ions at $T_b < 10^4 \, K$ gives rise to the formation of {\it 
molecular hydrogen} via the reactions~\footnote{As already pointed out, $H_2$ 
has no permanent electric dipole moment, and so the standard reaction $H + H 
\to H_2 + \gamma$ can only occur on dust grains. But since the composition of 
the protogalaxy is supposed to be primordial, no grains exist. Hence, the 
considered reaction is {\it not} 
operative. We stress that this fact implies that $CO$ is {\it not} a tracer 
of $H_2$ in the present context.} $H + p \to H^+_2 + \gamma$~, $H + e \to H^- 
+ \gamma$ and $H^+_2 + H \to H_2 + p$~, $H + H^- \to H_2 + e$ (notice that 
ions only act as catalysts). 

The presence of $H_2$ in the bubbles brings about a {\it further cooling}, 
owing to photon emission in roto-vibrational molecular transitions. We stress 
that this process is {\it very efficient} and can cause a temperature drop 
down to $T_b \sim 10 \, K$. Evidently in such a situation $t_{cool} \ll 
t_{ff}$, and so the bubbles collapse in free-fall and fragment. When the 
number density in 
the bubbles exceeds $\sim 10^8 \, cm^{- 3}$ -- which corresponds to 
$\rho_b \simeq 
2.8 \cdot 10^6 \, M_{\odot} \, pc^{- 3}$ -- a further $H_2$ production takes 
place via the three-body reactions $H + H + H \to H_2 + H$ and 
$H + H + H_2 \to H_2 
+ H_2$, as realized by Palla, Salpeter and Stahler~(1983). At variance with 
the former reactions, the latter ones do {\it not} 
involve ions as catalysts. As a consequence, the bubbles get transformed 
almost entirely into $H_2$. Besides modifying their internal composition, 
this circumstance makes cooling even more efficient.

Still, {\it a priori} nothing ensures that $H_2$ -- once produced -- will 
survive: because of its fragility, its actual existence crucially depends on 
the environmental conditions.

In fact, in the central region of the protogalaxy an $AGN$ (Active Galactic 
Nucleus) along with a first population of massive stars (population $III$) 
are expected to form. They act as strong sources of ultraviolet radiation, 
which dissociates the $H_2$ molecules up to a critical galactocentric distance 
$r_*$ (because the radiative flux decreases as the galactocentric distance 
increases). It is not difficult to estimates that the $H_2$ destruction 
should occur for $r < r_*$, with $r_* = 10 - 20 \, kpc$. Consequently, the 
further evolution of the {\it inner} region of the protogalaxy ($r < 10 - 20 
\, kpc$) will be {\it totally different} from that of the {\it outer} part 
($r > 10 - 20 \, kpc$). Actually, since the spheroid of radius $10 - 20 \, 
kpc$ gives rise to the inner halo of a normal spiral galaxy, we will 
distinguish between {\it inner} and {\it outer} halo.  

Owing to the above considerations, we are now in position to summarize the 
dynamics of the halo as follows.

\begin{itemize}

\item {\it Inner halo}~(Fall \& Rees 1985) -- The lack of $H_2$ prevents the bubbles from cooling 
down to temperatures smaller than $\sim 10^4 \, K$, and so they stay for a 
long time in hydrostatic equilibrium, which coincides with virial equilibrium 
(since $t_{cool} \gg t_{ff}$ for $T_b < 10^4 \, K$). As a result, their mass 
and radius are given (respectively) by the corresponding Jeans values 
(\ref{v10}) and (\ref{v11}). Indeed, the main result of the Fall-Rees theory 
is that these values are in {\it agreement} with those observed for globular 
clusters provided that mass loss during formation and evolution is taken into 
account. Subsequently the ultraviolet flux decreases, thereby allowing 
for the formation and survival of $H_2$. Accordingly, the bubbles can further 
cool, collapse in free-fall and fragment, ultimately producing ordinary 
stars. Thus, the Fall-Rees theory provides a natural explanation for the 
formation of {\it globular clusters} in the halos of normal spiral galaxies.

\item {\it Outer halo}~(De Paolis et al. 1995a,b) -- The presence of $H_2$ 
allows the bubbles to cool 
down to temperatures {\it much lower} than $10^4 \, K$, thereby collapsing in 
free-fall and fragmenting. As is well known, the collapse 
becomes adiabatic -- and eventually stops -- when a fragment becomes {\it 
optically thick}. Palla, Salpeter and Stahler~(1983) have shown that this 
occurs 
when the fragment Jeans mass is as low as $10^{- 2} - 10^{- 1} \, M_{\odot}$. 
So, clusters of {\it brown dwarfs} in the mass range $10^{- 2} - 10^{- 1} \, 
M_{\odot}$ should form in the outer halos of normal spiral galaxies. 

\end{itemize}
Notice that the present dark matter model gives rise to 
an initial mass function in the halo that {\it depends} on the galactocentric 
distance~\footnote{Evidence for a spatially varying initial mass function in 
the Milky Way disk has been reported by Taylor (1998).}~\footnote{Because of 
this fact, no extrapolation from the initial mass function in the disk can 
yield informations about the initial mass function in the halo. Consequently, 
the bound on the fraction of halo brown dwarfs derived by Graff and Freese 
(1996a) is {\it not} valid in the present context.}.

As a matter of fact, recent N-body simulations of hierarchical clustering 
produce dark matter halos with a universal {\it Navarro-Frenk-White} (1997) 
(NFW) density profile. However, this profile gets modified by the baryonic 
infall (associated with the disk formation)~(Blumenthal et al. 1986; Mo, Mao 
\& White 1998). In the 
case of Milky-Way-type galaxies, over galactocentric distances ranging from a 
few $kpc$ up to nearly $50 \, kpc$ the NFW density profile is turned into a 
SIS profile, which gives rise to the observed flat rotation curves. 
Manifestly, this result contradicts the Fall-Rees assumption of a SIS profile 
{\it before} baryonic infall, and so invalidates the above discussion. For 
instance, eqs. (\ref{v1}) and (\ref{v7}) imply that the fractional abundance 
of baryonic matter {\it increases linearly} with the galactocentric distance 
and should eventually dominate the mass density, but numerical 
experiments disprove this behaviour. In addition, self-consistency of the 
model (${\rho}_d < {\rho}_o$) demands that the radius of the dark halos 
should obey the constraint $r < 83 \, {\alpha}^{ - 1/2} \, kpc$, but again 
there is good observational evidence that dark halos are considerably more 
extended (Zaritsky et al. 1997)~\footnote{Besides, it has become clear that 
the formation of globular clusters does not proceed in accordance with the 
Fall-Rees theory (see e.g. Fall \& Zhang 2001; Van den Bergh 2001; Beasley et 
al. 2002).}. 

Yet, 
the main point of the present dark matter scenario -- namely the {\it survival 
of $H_2$ at large galactocentric distances}, which gives rise to the dark 
cluster formation via efficient cooling and fragmentation -- is {\it 
independent} of the specific halo density profile. So, the {\it qualitative 
features} of the model in question are expected to remain true even within the 
contemporary picture of galaxy formation. Unfortunately, before this issue 
is settled no prediction about the actual distribution of the dark 
clusters in galactic halos can be made.

\subsection{Properties of the dark clusters}

Let us summarize the main properties of the dark clusters, as suggested by 
the above considerations.

Dark clusters resemble both morphologically and dynamically globular 
clusters, apart from the obvious difference of being made of brown dwarfs 
rather than ordinary stars. This fact entails in turn further, crucial 
differences, which we are now going to discuss.

A difference between globular and dark clusters concerns their mass 
spectrum. In fact, globular cluster masses exhibit a small scatter around a 
preferred value $\sim 10^5 \, M_{\odot}$. Within the Fall-Rees model, this 
circumstance is naturally traced to the formation mechanism. Indeed, the 
permanence 
of the corresponding bubbles for a long time in virial equilibrium results in 
the imprinting of the associated Jeans mass, which is just $\sim 10^5 \, 
M_{\odot}$ when eq. (\ref{v10}) is corrected for the mass loss. On the other 
hand, the bubbles that generate the dark clusters cool {\it monotonically}, 
and so {\it no} preferred mass scale gets singled out. Accordingly, we expect 
the dark cluster mass spectrum to be much {\it wider}. However, disruptive 
effects -- like evaporation, encounters (both among themselves and with 
globular clusters), tidal disruption and spiralling motion towards the 
galactic centre (brought about dynamical friction) -- strongly constraint 
their mass range~\footnote{A general analysis of these constraints has been 
carried out by Carr and Lacey (1987), Moore and Silk (1995), and Carr and 
Sakellariadou (1998). As they assumed that the dark cluster distribution 
extends all the way down to the galactic centre, their results do {\it 
not} apply to the present model.}. Specifically, {\it only} dark clusters in 
the mass range $3 \cdot 10^2 - 10^6 \, M_{\odot}$ are expected to survive all 
these effects, and therefore to still populate the outer galactic halos {\it 
today}~(De Paolis et al. 1996, 1998a). We stress that this is possible {\it 
only} because the dark matter model in question predicts that dark clusters do 
{\it not} populate the inner halo. Typical values of the 
dark cluster radius lie in the range $1 - 10 \, pc$. Moreover, all dark 
clusters more massive than $\sim 10^{5} M_{\odot}$ have presumably entered the 
phase of core collapse.

Another difference between globular and dark clusters concerns their gas 
content. Because the process of star formation is highly {\it inefficient}, 
at least 
$60\%$ of the original amount of gas does {\it not} get transformed into stars
~(Scalo 1985). Within globular clusters, such a leftover gas is expelled by 
stellar winds and shock waves driven by supernova explosions. The case of 
dark clusters is different. Since practically no nuclear process and no 
evolution occur in the brown dwarfs, the leftover gas -- which is mainly 
$H_2$ -- remains trapped inside the dark clusters in the form of 
self-gravitating {\it clouds}. Indeed, we expect these gas clouds to 
provide the {\it leading} mass contribution to the dark clusters.

Although these clouds are primarily made of $H_2$, they are expected to be 
surrounded by a layer of {\it neutral hydrogen} $HI$ and a {\it ionized 
``skin''}, owing to the interaction with the diffuse photon background~(De 
Paolis et al. 1996, 1998a). As we shall see, the presence of $HI$ and ions in 
the outer region of the clouds plays an important r\^ole in some subsequent 
considerations.

An analysis of the cloud thermal balance shows that they are very cold, 
with a central temperature $\sim 10 \, K$. Moreover, Gerhard and Silk
~(1996) have demonstrated that for realistic values of their parameters the 
clouds should survive evaporation and collisional disruption. Typical values 
of the cloud mass and radius are $\sim 10^{- 3} \, M_{\odot}$ and 
$\sim 10^{- 5} \, pc$, respectively. 

Mainly in view of the observational implications, an important question 
concerns the {\it present} dust-to-gas ratio of the clouds. Unfortunately, a 
clear-cut answer cannot be given. It looks natural to suppose that the clouds 
formed from almost primordial gas. But even if they were essentially dust free 
at the beginning, their interaction with the interstellar matter -- when they 
periodically pass through the galactic disk -- should have produced a sizable 
dust-to-gas ratio today, as pointed out by Kerins, Binney and Silk (2002). 
Accordingly, 
it has been suggested that the clouds should be {\it opaque} to visible light 
(Gerhard \& Silk 1996; Kerins, Binney \& Silk 2002). However, this conclusion 
may not 
be true. For instance, Draine (1998) has argued that dust grains could 
have sedimented in a small core, so that a cloud would be essentially {\it 
transparent} at optical wavelenghts. In the following, both possibilities 
will be considered.

Cloud stability is a critical issue. Although a completely satisfactory 
treatment is still lacking, various mechanisms have been envisaged that 
can stabilize the clouds against gravitational collapse~\footnote{As we will 
see later, the clouds are likely to be magnetized. However, it is not clear 
whether this fact helps to stabilize them, since the energy balance depends 
critically on the configuration of the magnetic field.}. A possibility 
investigated by Gerhard and Silk (1996) is that the clouds get {\it 
stabilized} by the gravitational field produced by the brown dwarfs clumped 
into the dark clusters. Alternatively, it has been suggested that cosmic-ray 
heating can balance either molecular cooling in dust-free clouds (Sciama 
2000) or cooling by dust emission in dusty clouds (Lawrence 2001). Finally, 
Wardle and Walker (1999) have argued that stability can be achieved via 
sublimation of liquid or solid hydrogen, which is expected to be present in 
clouds with mass $< 0.02 \, M_{\odot}$.

In turn, it looks remarkable that the presence of gas clouds in the dark 
clusters can affect the properties of the brown dwarfs. Indeed -- as shown by 
Hansen~(1999) -- the cloud-brown dwarf interaction can give rise to  a 
low-entropy accretion process~\footnote{We stress that this process is likely 
to occur in the quiet environment inside the dark clusters.}~(Lenzuni, 
Chernoff \& Salpeter 1992). Accordingly, the brown 
dwarf mass gets substantially increased -- up to $0.3 \, M_{\odot}$ -- while 
the central temperature stays below the hydrogen-burning threshold. Although 
it would then be more appropriate to talk about {\it beige dwarfs} -- as 
indeed suggested by Hansen -- we will continue to use the name {\it brown 
dwarfs} for simplicity, but it should be kept in mind that {\it their mass 
can be as large as $0.3 \, M_{\odot}$}.

Finally, we point out that also {\it binary} brown dwarfs should be 
clumped into the dark clusters. In fact -- much in the 
same way as it occurs for ordinary stars -- also in this case the 
fragmentation process is expected to produce a large fraction (up to $50\%$ 
in mass) of binary objects~\footnote{Binary brown dwarfs tend to concentrate 
in the dark cluster cores owing to the mass-stratification 
instability~(Spitzer 1987).}. 
Because of dynamical friction on the gas clouds, the overwhelming majority 
of binary brown dwarfs become so close (hard) today that they cannot be 
resolved in current gravitational microlensing experiments towards the 
Magellanic Clouds (more about this, later)~\footnote{We remark that the 
energy released during the hardening process of binary brown dwarfs is 
efficiently radiated away by the clouds.} (De Paolis et al. 1998b).

\section{Observational implications} 

The formation mechanism discussed in Subsect. 2.1 obviously provides a 
rationale for the existence of dark clusters with the properties stated in 
Subsect. 2.2. Yet, a more pragmatic attitude can also be taken (Gerhard \& 
Silk 1996). Indeed -- regardless of the specific formation process -- one can 
simply {\it suppose} 
that dark clusters of brown dwarfs~\footnote{Recall that by brown dwarfs we 
actually mean beige dwarfs throughout the paper.} and cold $H_2$ clouds account 
for baryonic dark matter and populate the halos of normal spiral galaxies at 
galactocentric distances $> 10 - 20 \, kpc$. 

What are then the observational 
signatures of this scenario? Below, we discuss a few most important effects.

${\gamma}$-{\it ray emission} -- At the qualitative level, the 
situation looks quite simple. Very high-energy ($E > 1 \, GeV$) cosmic-ray 
protons travelling in the halo of a normal spiral galaxy produce (in 
particular) 
{\it neutral} pions upon scattering on the gas clouds. And of course the pions 
rapidly decay into photons. So, a $\gamma$-ray emission is expected, whose 
flux 
is proportional to the colomn density of the clouds.

Unfortunately, a quantitative analysis is difficult, since nobody knows how 
the cosmic-ray protons propagate inside the dark halos~\footnote{We stress 
that -- contrary to the practice used in the cosmic-ray community -- by {\it 
halo} we mean the almost spherical galactic component which extends well 
beyond $\sim 10 \, kpc$, and not the {\it thick disk.}}. Hence, some further 
assumptions are definitely needed at this point.

A strategy to proceed can be sketched as follows, focusing the attention on 
the 
Milky Way~(De Paolis et al. 1999, 2000). First of all, recall that in the 
Milky Way {\it disk} cosmic rays are produced by stellar winds and 
supernova explosions, and tend to escape into intergalactic space. However -- 
owing to the inhomogeneities of the disk magnetic field over scales $10^{-6} 
- 10^2 \, pc$ -- they undergo a diffusion process, which gives rise to a 
temporary confinement controlled by the escape time from the disk. Next, 
observe that -- within the present dark matter model -- a {\it similar} 
situation is expected to occur in the Milky Way halo as well. For, we have 
seen that the gas clouds -- with a 
photo-ionized ``skin'' -- have typical size $\sim 10^{- 5} \, pc$, and they 
are clumped into dark clusters having typical size $\sim 10 \, pc$. Thus, we 
expect inhomogeneities of the halo magnetic field of the {\it same kind} as 
the ones occurring in the disk~\footnote{This picture is supported by 
Kronberg's (1994) suggested existence of a cosmic background magnetic field 
with strength $\sim 1 \, {\mu}G$, which is motivated by the fact that 
magnetic fields with this strength are found nearly everywhere, regardless 
of the actual density and composition of the corresponding region. Moreover, 
the inhomogeneities of the magnetic field inside a dark cluster are expected 
to be produced both by the ionized envelop of the clouds and by the turbolent 
dynamo process likely present in the brown dwarf coronae (Drake et al. 
1996).}. This circumstance allows for an estimate of 
the cosmic-ray escape time from the halo. It turns out that cosmic-ray protons 
with energy $E < 10^3 \, GeV$ have an escape time {\it larger} than the Hubble 
time. Still, their spectrum goes like $E^{- \alpha} \, (2 < \alpha 
< 3)$, and so just these protons give the {\it leading} contribution to 
the cosmic-ray proton flux in the halo. Therefore, most of the cosmic-ray 
protons produced in the Milky Way disk should still be trapped inside 
the Galactic halo {\it today}. As a result, we can estimate the average 
density of cosmic-ray protons in the 
halo, which turns out to be~\footnote{This value is consistent with the upper 
bound derived from {\it EGRET} observations (Sreekumar et al. 1993).} 
$\sim 0.1 \, eV \, cm^{- 3}$, namely roughly 
one-tenth of the disk value. Such a high cosmic-ray proton flux gives rise to 
a potentially detectable {\it $\gamma$-ray flux} from the Milky Way halo 
through 
the above-mentioned mechanism. Owing to the very poor angular resolution of 
present-day $\gamma$-ray detectors, this flux is {\it indistinguishable} 
from a 
truly diffuse emission from the Galactic halo. Although its intensity depends 
on various somewhat uncertain parameters, an order-of-magnitude estimate 
yields for the integrated flux above 1 GeV~(De Paolis et al. 1995a,b)
\begin{equation}
{\Phi}_{\gamma}(> 1 \, GeV) = 10^{-7} - 10^{-6} \, \, \, \, {\gamma} \, 
cm^{-2} \, s^{-1} \, sr^{-1}~.  \label{r11}  
\end{equation}
We stress that an independent calculation~(Kalberla, Shchekinov \& Dettmar 
1999) -- based on different assumptions -- leads to the same conclusion.

In spite of the fact that the flux in eq. (\ref{r11}) is slightly less than 
the 
observed diffuse extragalactic flux~(Sreekumar et al. 1998), a wavelet-based 
statistical analysis can {\it discriminate} between a Milky Way halo emission 
and an extragalactic flux. In $1998$ such an analysis has been carried out for 
{\it EGRET} data and has led to the discovery of a diffuse $\gamma$-ray 
emission from the 
Galactic halo~(Dixon et al. 1998). Remarkably enough, the observed flux is in 
agreement with both the intensity and the spatial distribution of the emission 
predicted by the considered dark matter model, provided that a moderate halo 
flattening is allowed~\footnote{As already pointed out, the dark matter model 
in question does not predict (in its present form) how dark clusters are 
distributed in the Galactic halo. While it looks natural to suppose that 
their distribution follows that of the nonbaryonic dark matter -- namely a 
SIS density profile -- some 
flattening can be expected~(Samurovic, Cirkovic \& Milosevic-Zdjelar 1999). 
In the present calculation, the {\it baryonic halo} is 
modelled as a {\it flattened spheroid} -- which becomes a SIS in the limit of 
spherical symmetry -- with the flattening parameter left free. Because 
nonbaryonic dark matter dominates over the baryons, a moderately flattened 
baryonic dark halo is consistent with the recent evidence of a nearly 
spherical Galactic halo (Ibata et al. 2001).}~(De Paolis et al. 1999, 2000). 

It goes without saying that the future planned satellite missions {\it AGILE} 
and {\it GLAST} will play a crucial r\^ole in settling this issue. 

As is well known, the Milky Way is a {\it typical} normal spiral galaxy, and 
so we 
expect such a $\gamma$-ray halo emission from nearly all normal spiral 
galaxies. In particular, observing the $\gamma$-ray flux from the halo of 
Andromeda galaxy will be a challange for the next generation $\gamma$-ray 
detectors~(De Paolis et al. 2000).

A final comment is in order. When high-energy cosmic-ray protons scatter on 
the gas clouds, also {\it charged} pions are produced, which ultimately give 
rise to a flux of high-energy electrons~\footnote{Positrons disappear because 
of annihilation into photons, thereby increasing the previously-considered 
$\gamma$-ray flux.}~\footnote{Also high-energy neutrinos arise in this 
process, but a preliminary analysis~(Ingelman \& Thunman 1996) has shown that 
this flux does not compete with that of neutrinos produced in the Milky Way 
disk.}. Were these electrons to escape from a cloud, two further 
effects would come about and should be addressed. One is emission of 
synchrotron radiation in the halo magnetic field. The other is inverse 
Compton scattering against CMBR photons, leading to a soft X-ray flux. 
Although in either case the intensity is expected to be {\it 
subdominant}~\footnote{The $\gamma$-ray flux discovered by Dixon et al. 
(1998) is slightly smaller than the diffuse extragalactic flux, which is in 
turn considerably smaller than the flux from the Milky Way disk (Hunter et 
al. 1997). Because the branching ratio is obviously independent of the 
environment, also the electron flux from the halo should be considerably 
smaller than the corresponding electron flux from the Milky Way disk.}, the 
latter effect might be disentangled from the background by its characteristic 
angular distribution. However, the relativistic electrons hardly escape from 
a cloud -- owing to its ionized outer envelope -- and so these effects fail 
to provide a signature of the dark matter scenario under consideration.

{\it X-ray emission} -- As already emphasized, brown dwarfs are expected to 
possess a coronal soft X-ray emission. More explicitly, Kayshap et al. (1994) 
have quantified this flux in $\sim 10^{27} \, erg \, s^{- 1}$ in the $0.1 - 
10 \, keV$ energy range. Consequently, the possibility arises to discover the 
dark clusters in X-ray searches.

A thorough analysis has shown that this should in fact be the case for dark 
clusters with mass as large as $10^{5} \, M_{\odot}$, provided that the 
fractional abundance of brown dwarfs in the dark clusters is sizable~(De 
Paolis et al. 1998c). Unfortunately, it is impossible to tell {\it a priori} 
whether this circumstance is indeed realized, and so an intrinsic uncertainty 
affects the present discussion. More specifically, dark clusters with mass 
$\sim 10^{5}$ $M_{\odot}$ can contribute to the new population of faint X-ray 
sources advocated by Hasinger et al.~(1993) and by McHardy et al.~(1997), 
whereas dark clusters with mass $\sim 10^{6}$ $M_{\odot}$ can be observed 
as resolved sources with the future planned satellite missions.

We stress that these results were obtained in 1998, when only {\it ROSAT} 
data were available. The impact of the {\it Chandra} and {\it XMM-Newton} 
missions on this issue has still to be investigated. 

{\it Infrared emission} -- A different kind of dark matter search 
addresses the infrared emission 
from very low-mass stars in the dark halos of various galaxies, including the 
Milky Way. Although strong constrains have been set on the abundance of red 
dwarfs, dark halos dominated by {\it old} brown dwarfs are still a viable 
possibility~(Boughn \& Uson 1995; Graff \& Freese 1996b; Gilmore \& Unavane 
1998). In this respect, it should be 
stressed that all the analyses performed so far rest upon the strong 
assumption of a {\it smooth} distribution of low-mass stars within a {\it 
canonical} SIS halo model~\footnote{Hence, low-mass stars are spread out all 
the way down to the galactic centre with a $r^{- 2}$ density profile.}. Yet, 
we have seen that within the present dark matter scenario things are 
different. 
Brown dwarfs are clumped into the dark clusters, and this very fact {\it 
reduces} the expected infrared flux~(Kerins 1997a,b). In addition, we expect 
gas 
clouds -- rather than brown dwarfs -- to give the leading mass contribution 
to the dark clusters, and so the resulting infrared flux gets {\it further 
reduced}~\footnote{In the case of the Milky Way, yet another reduction of the 
infrared flux is due to the fact that dark clusters populate {\it only} the 
outer halo.}. Therefore only future observations might detect the infrared 
emission in question.

Infrared emission from both dust-less (Sciama 2000) and dusty (Lawrence 2001) 
clouds has been considered in a different context, with the conclusion that 
only for clouds close to the Milky Way disk can the resulting flux lie above 
the threshold of the infrared detector {\it SCUBA}.

{\it Occultation effects} -- As discussed in Subsect. 2.2, there are good 
reasons to expect that the gas clouds clumped into the dark clusters are 
{\it opaque} 
at optical wavelenghts. Consequently, they can be detected by looking for 
occultations of background stars (Gerhard \& Silk 1996). Quite recently, 
Kerins, Binney and Silk (2002) have suggested that the data sets of 
gravitational 
microlensing experiments towards the Magellanic Clouds (see below), the 
Galactic bulge and the Andromeda galaxy can also be used to search for 
occultation signatures by gas clouds. Actually, they have demonstrated that 
-- for cloud parameters typical of the considered dark matter model -- 
thousands of transit events should already exist within microlensing survey 
data sets. 

{\it Optical lensing} -- It might nevertheless happen that the gas clouds are 
effectively {\it transparent} at optical light. In this case, the light of a 
background 
star can be magnified -- in a symmetric fashion -- when a cloud crosses its 
line of sight. The resulting light curves resemble those of gravitational 
microlensing (see below), apart from the fact that red light suffers a {\it 
stronger} magnification than blu light (because of Rayleigh scattering). The 
phenomenology of the corresponding {\it chromatic} events has been studied by 
Draine (1998) and by Rafikov and Draine (2001), who have computed the event 
rates in the case of background stars in the Large Magellanic Cloud (LMC). 
They also suggested that searches for gravitational microlensing could be 
used to detect optical lensing events, provided that the achromaticity 
constraint imposed so far is relaxed (a study of the absorption features 
caused by the gas clouds would help to identify the desired events).

{\it Gravitational microlensing} -- This effect~\footnote{A thorough account 
of microlensing can be found in Mollerach and Roulet (2002).} towards the 
Magellanic Clouds was proposed by Paczynski ~(1986) as a tool to 
discover {\it compact dark matter objects} -- named MACHOs (Massive 
Astrophysical Compact Halo Objects) -- presumably lurking in the halo of the 
Milky Way. Basically, a background star gets magnified -- in a symmetric and 
achromatic fashion -- when a MACHO crosses its line of sight. Since $1993$, 
the {\it MACHO} collaboration has detected $13 - 17$ 
events towards the LMC, the observed optical depth 
being ${\tau}_{obs} \simeq 1.2 \cdot 10^{ - 7}$~(Alcock et al. 2000). Only 4 
events have been 
found by the EROS2 collaboration~(Milsztajn \& Lasserre 2001)~\footnote{A 
comparison of the results of the {\it MACHO} and {\it EROS2} callaborations 
is not 
straightforward, since they look at different fields on the LMC, monitoring a 
different number of stars with different observation times.}. In the 
following, we shall focus our attention on the {\it MACHO} data. 

Because a $100 
\, \%$ MACHO {\it canonical} SIS halo model~\footnote{Accordingly, MACHOs are 
smoothly spread out all the way down to the Galactic centre with a $r^{- 2}$ 
density profile. However, it should be stressed that this 
model is purely academic, since it both assumes  a $100 \, \%$ efficiency in 
the MACHO formation process and neglects nonbaryonic dark matter.} 
predicts ${\tau}_{pred} \simeq 5 \cdot 10^{- 7}$, one would conclude that 
roughly $20 \, \%$ of the halo dark matter should be in the form of MACHOs. As 
the theory of galaxy formation requires most of the galactic dark matter to 
be nonbaryonic, this result looks reasonable, but we 
emphasize that it relies upon two strong assumptions: {\it all} microlensing 
events are due to MACHOs, and a {\it canonical} SIS MACHO distribution. As a 
matter of fact, 
the former assumption is grossly violated. Besides faint stars in the various 
components of the Milky Way (thin disk, thick disk, spheroid and halo), also 
faint stars insides the LMC itself can produce microlensing events~\footnote{
An updated discussion of the various contributions is given by Jetzer, 
Mancini and Scarpetta (2002).} (the latter 
phenomenon being referred to as {\it self-lensing}), but -- as a rule -- 
observations provide no information about the location of the lenses (unless 
they happen to be binary objects). Although various attempts at estimating 
the optical depth for self-lensing ${\tau}_{self}$ have been made, discordant 
results emerged (due to the poor knowledge of the mass distribution in the 
LMC), ranging from 
${\tau}_{self} = (0.7 - 7.8) \cdot 10^{- 8}$ (Gyuk, Dalal \& Griest 2000) up to 
${\tau}_{self} = (0.7 - 1.8) \cdot 10^{- 7}$ (Evans \& Kerins 2000)~\footnote{
In spite of the fact that the large value ${\tau}_{self} \simeq 1.4 \cdot 
10^{- 7}$ estimated by Weinberg (2000) has been criticized by Gyuk, Dalal and 
Griest (2000), similar large values have been obtained on different grounds 
by Evans and Kerins (2000) and Zhao and Evans (2000). Hopefully, the new data 
of Van der Marel et al. (2002) on the LMC will clarify the situation.}. 
Evidently no 
sharp conclusion can be drawn about the fractional abundance of MACHOs in the 
Milky Way halo, but it seems fair to state that a MACHO contribution to the 
optical depth ${\tau}_{MACHO} \sim 10^{ - 8}$ is quite consistent with 
present microlensing data. 

A set of microlensing events is also characterized by the {\it average} lens 
mass $\overline M$, which is related to the event duration and depends 
strongly on the assumed galactic model. The MACHO collaboration finds 
$\overline M \simeq 0.7 \, M_{\odot}$ for a {\it canonical} SIS halo model, 
whereas 
the extreme cases of a maximal and minimal halo yield $\overline M \simeq 0.9 
\, M_{\odot}$ and $\overline M \simeq 0.5 \, M_{\odot}$, respectively (with 
large uncertainties). Unfortunately, $\overline M$ does not possess any clear 
physical meaning, given that at least 5 different lens populations contribute 
to the optical depth (as already emphasized). Only by a detailed modelling 
of every population can the average mass corresponding to each population be 
estimated. According to Jetzer, Mancini and Scarpetta (2002), the MACHO 
average mass turns 
out to be ${\overline M}_{MACHO} \simeq 0.3 \, M_{\odot}$ for a {\it 
canonical} SIS halo model. Obviously -- within such a model -- ordinary 
brown dwarfs are ruled out as a canditate for MACHOs. 

Let us now discuss how the observed microlensing events fit within the 
considered dark matter scenario. It goes without saying that -- while not 
logically compelling -- here MACHOs are naturally identified with {\it brown 
dwarfs}, and this fact raises two questions which will be addressed separately.

\begin{itemize}

\item A question arises because then MACHOs make up a nonnegligible fraction 
of the halo dark matter, and so the {\it predicted} MACHO contribution to the 
optical depth ${\tau}'_{MACHO}$ might well come out {\it too large} (i. e. 
larger than $\sim 10^{- 8}$).

In order to clarify this issue, we compare the dark matter 
model in question with a {\it realistic canonical} SIS halo model. Because 
leftover gas as well as nonbaryonic dark matter have to  be taken into 
account, the MACHO fractional abundance in a realistic canonical SIS halo 
model cannot exceed, say, $20 \, \%$. Then the above considerations 
entail that the MACHO contribution to the optical depth {\it predicted} by a 
realistic canonical SIS halo model is ${\tau}''_{MACHO} \simeq 1.2 \cdot 
10^{ - 7}$. So, the critical point is whether the morphological difference 
between the two 
models indeed entails that ${\tau}'_{MACHO}$ is lower that ${\tau}''_{MACHO}$ 
by roughly one order of magnitude. We believe that the answer is yes, as we 
are going to show. 

A natural expectation is that also the baryonic dark halo under consideration 
has a SIS density profile -- but a {\it noncanonical} one -- since {\it only} 
the outer halo is populated by dark clusters, hence by MACHOs. Thus, we can 
formally regard the present dark matter model as a realistic canonical SIS 
model with {\it no} MACHOs in the inner part. Consequently, 
${\tau}'_{MACHO}$ is reduced with respect to ${\tau}''_{MACHO}$ and we 
actually get~\footnote{The uncertainty depends on the minimal galactocentric 
distance of the dark clusters $r = 10 - 20 \, kpc$.} ${\tau}'_{MACHO} \simeq 
( 0.4 - 0.8 ) \cdot 10^{- 7}$~(Jetzer 2001). Notice that a moderate flattening 
of the baryonic dark halo -- as suggested by the observed ${\gamma}$-ray 
emission -- would further lower 
${\tau}'_{MACHO}$. As a matter of fact, we expect the MACHO fraction to be 
{\it less} than $20 \, \%$, because the dark clusters should mainly consist 
of $H_2$ clouds -- instead of brown dwarfs -- and this circumstance causes an 
additional reduction of ${\tau}'_{MACHO}$. Gas clouds are not compact 
enough to produce microlensing events (Henriksen \& Widrow 1995), but -- 
depending on their dust-to-gas ratio -- can either obscurate or magnify a 
background star. As pointed out above, in the latter case the effect is 
chromatic. Therefore -- owing to the achromaticity constraint imposed in 
current microlensing searches -- the resulting events would simply be 
discarded. Now, within this context MACHOs are associated with the gas 
clouds, and so some would-be genuine microlensing events -- produced by 
MACHOs -- can either become chromatic or disappear altogheter owing to an 
intervening gas cloud~\footnote{A quantitative analysis of these chromatic 
microlensing events has been performed by Bozza et al. (2002).}. Because the 
resulting events would {\it not} be observed in present-day 
experiments, yet another reduction of ${\tau}'_{MACHO}$ comes about. On the 
whole, there should be little doubt that the desired reduction can indeed be 
achieved, even if a precise estimate would be impossible.

An alternative possibility has been suggested by Kerins and Evans (1998), who 
suppose that the initial mass function in the halo varies {\it smoothly} with 
the galactocentric distance. Observe that here the baryonic halo model 
necessarily differs from the SIS. They have shown that in such a situation 
brown dwarfs {\it fail} to dominate the optical depth while still dominating 
the (baryonic) mass density. As a result, the risk of too large values of 
${\tau}'_{MACHO}$ disappears~\footnote{Because it is not clear at this stage 
whether MACHOs are really brown dwarfs in this context (see below), here 
${\tau}'_{MACHO}$ actually denotes the brown dwarf contribution to the 
optical depth.}.

\item Another question concerns the predicted value of ${\overline M}_
{MACHO}$, because it can turn out to {\it largely exceed} the brown dwarf 
mass, thereby preventing MACHOs from being brown dwarfs.

Again, let us consider first a (noncanonical) SIS halo model. An 
explicit calculation shows that the lack of brown dwarfs in the inner halo 
produces a reduction of ${\overline M}_{MACHO}$, and we get 
${\overline M}_{MACHO} \simeq 0.2 \, M_{\odot}$~(Jetzer 2001). As before, a 
moderate flattening of the baryonic dark halo further lowers 
${\overline M}_{MACHO}$. 
Yet, we have seen that within the considered dark matter model brown dwarfs 
are to be replaced 
by {\it beige dwarfs}, with masses up to $0.3 \, M_{\odot}$. Moreover, a 
substantial fraction of these beige dwarfs should be binary systems, which 
are presently so close (hard) that {\it cannot} be resolved in current 
microlensing searches (De Paolis et al. 1998b). As a result, the {\it 
effective brown dwarf mass} can well be as large as 
$0.3 \, M_{\odot}$. So, no problem arises.

Within the Kerins-Evans scenario the situation is different and depends on 
the quantitative details of the model. However, ${\overline M}_{MACHO}$ 
tends to come out close to the value predicted by the canonical SIS model. 
Therefore MACHOs are {\it not} brown dwarfs in this setting, and their nature 
is an oper question. A possibility could be that MACHOs are white 
dwarfs~(Oppenheimer et al. 2001). Because of their low fractional abundance, 
the constraints discussed in Sect. 1 might be circumvented.

\end{itemize}

\noindent In conclusion, the considered dark matter model is fully 
consistent with present-day microlensing data~\footnote{Because so few 
microlensing events are produced by MACHOs, it is impossible to decide 
observationally whether or not MACHOs are clustered~(Maoz 1994; Metcalf \& 
Silk 1996).}. 

It goes without saying that detection of {\it chromatic} microlensing events 
in future experiments would provide evidence that MACHOs are indeed surrounded 
by gas clouds.

{\it CMBR anisotropies} -- Dark clusters can also be discovered by performing 
high-precision measurements of the CMBR anisotropy. This strategy involves a 
sort of {\it kinematic Sunyaev-Zeldovich effect}~(Sunyaev \& Zeldovich 1980) 
in the gas clouds clumped into the dark clusters and can be illustrated as 
follows~\footnote{We are neglecting here the kinematic Sunyaev-Zeldovich 
effect arising from the Compton scattering of CMBR photons on the ionized 
``skin'' of the clouds.}~(De Paolis et al. 1995c). Assuming for 
simplicity that dust effects can be neglected, absorption and emission 
processes merely involve molecular roto-vibrational transition lines, 
because of the very low cloud temperature. So, photons of the CMBR can be 
absorbed and re-emitted by the clouds. Since the latter photons are evidently 
Doppler-shifted -- owing both to the cloud velocity dispersion inside the 
cluster and to the cluster peculiar velocity -- an anisotropy in the CMBR 
shows up when looking towards a dark cluster.

Of course, in order for the effect to be sizable a sufficiently large number 
of photons must be involved, and the transition line in question has to be 
optically thick. Hence, the real question is whether any such molecular line 
falls close enough to the pick of the CMBR. Although the answer obviously 
depends on the poorly-known cloud composition, even in the limiting case of a 
primordial metallicity the first rotational transition line of $LiH$ lies 
very close to the CMBR pick and is optically thick~\footnote{This is 
basically due to the fact that the very low colomn density is compensated 
by the huge resonance cross-section.}. 

Observe that for dark clusters in the Milky Way halo the resulting CMBR 
anisotropies show up on the angular scale of $\sim 1 \, arcminute$. This 
happens to be just the typical angular scale associated with CMBR 
anisotropies produced -- via the Sunyaev-Zeldovich effect -- by clusters of 
galaxies (Rephaeli 1995; Birkenshow 1999). Consequently, the same observing 
strategies employed in the latter case (for a review, see Rephaeli 2001) can 
be used to detect the dark clusters in the Milky Way halo. Obviously, only 
the occurrence of a microlensing event can tell us the actual position of a 
dark cluster on the sky -- in real time, if microlensing data are analyzed 
on-line -- and so a coordinated effort is required. 

A nonconventional procedure to detect dark clusters in the Andromeda halo -- 
which operates on the angular scale of $\sim 1 degree$ -- has been proposed 
by De Paolis et al. (1995c).

{\it Extreme Scattering Events} (ESEs) --  They are dramatic flux changes 
occurring -- over several weeks to months -- during radio flux monitoring of 
some quasars (Fiedler et al. 1987). It is generally agreed that ESEs are 
{\it not} intrinsic variations, but rather apparent flux 
changes caused by radio wave {\it refraction} when a cloud crosses the line 
of sight to a quasar. Evidently, in order to produce an ESE the cloud has to 
be {\it ionized}, and the radio signal features demand that the cloud radius 
and electron density should be $\sim 10^{- 5} \, pc$ and $\sim 10^3 \, 
cm^{- 3}$, respectively. Assuming {\it full ionization} may look natural at 
first sight, but then the resulting electron pressure turns out to exceed 
that of the interstellar medium by a factor $\sim 10^3$, thereby leading 
to complete evaporation of the cloud within $\sim 1 \, yr$. 

In $1998$, Walker 
and Wardle~(1998) claimed that the first consistent explanation of the ESEs 
involves an unclustered population of cold ($T \sim 10 \, K$) $H_2$ clouds 
with radius $\sim 10^{- 5} \, pc$ and mass $\sim 10^{- 3} \, M_{\odot}$, 
distributed in a canonical SIS halo and comprising most of its mass. 
Basically, radio way refraction is caused by the photo-ionized ``skin'', 
whereas the inner neutral region keeps the electron pressure sufficiently 
small. 

It looks intriguing that these clouds are practically identical to 
those clumped into the dark clusters, and so it would tempting to imagine 
that the present dark matter scenario could explain the ESEs. However, the 
cloud spatial distribution is very different in the two models and the effect 
of clustering on the ESEs has to be investigated before any conclusion can be 
drawn.

{\it Lyman-${\alpha}$ clouds} -- A typical line of sight to a distant quasar 
contains a huge number of absorption features, brought about by intervening 
gas clouds with colomn density in the range $3 \cdot 10^{12} - 2 \cdot 
10^{20} \, cm^{- 2}$. In particular, a {\it forest} of Lyman-${\alpha}$ 
absorption lines of $HI$ is associated with clouds having colomn density in the 
range $3 \cdot 10^{12} - 3 \cdot 10^{15} \, cm^{- 2}$. Because it is 
difficult to establish whether the Lyman-${\alpha}$ lines are produced by 
transient filamentary intergalactic gas structures or by gas clouds clustered 
around galaxies, the physical nature of Lyman-$\alpha$ absorption systems is 
still controversial. While both options are likely to be -- 
at least partially -- correct, Lanzetta and collaborators claim that {\it 
low-redshift} Lyman-${\alpha}$ lines are {\it mostly} 
produced by gas clouds inside dark galactic halos~(Chen et al. 1998, 2001). 
Moreover, Chen, Prochaska and Lanzetta (2001) have estimated the cosmological 
equivalent amount of $HI$ in these Lyman-${\alpha}$ clouds to be 
${\Omega}_{HI,g} = (2 - 8) \cdot 10^{- 3}$.

Naturally, the question arises as to whether the clouds clumped into the dark 
clusters can be {\it identified} with the low-redshift Lyman-${\alpha}$ clouds. 
Superficially, the answer seems to lie in the negative. For, we have seen 
that typical cloud parameters are $M \sim 10^{- 3} \, M_{\odot}$ and $R \sim 
10^{- 5} \, pc$, resulting in an average colomn density of $\sim 10^{25} \, 
cm^{- 2}$, 
namely ten orders of magnitude larger than allowed. However, such a colomn 
density actually pertains to $H_2$, {\it not} to $HI$. In fact, two points 
should be emphasized.

\begin{itemize}

\item Given that $HI$ is present only in the outer region of the 
clouds~\footnote{Because of this fact, the effect of dust should be 
irrelevant here, since dust is expected to dominate the inner part of a 
cloud.}, its volume density is presumably {\it much smaller} than the average 
cloud density.

\item The reported typical values of the Lyman-${\alpha}$ cloud colomn density 
are inferred under the implicit assumption of {\it full} $HI$ cloud 
composition. 
However, this is presently {\it not} the case, and what matters here is the 
{\it thickness} of the external $HI$ layer (rather than the cloud radius), 
which is expected to be {\it several orders of magnitude smaller} than the 
cloud size. 
\end{itemize}
Altogether, the actual $HI$ colomn density can well lie in the allowed range. 
Furthermore, the fractional abundance of $HI$ appears to fit within the above 
estimate if the clouds in questions indeed provide the main contribution to 
the dark baryon budget at low-redshift. Needless to say, only a detailed 
analysis of the cloud distribution predicted by the considered dark matter 
model can show whether the observed properties of low-redshift 
Lyman-${\alpha}$ forest lines are correctly reproduced.

{\it $H_2$ absorption lines} -- No doubt, in principle the most 
straigthforward way to discover the clouds clumped into the dark clusters is 
based upon detection of ultraviolet Lyman and Werner absorption lines of 
$H_2$ in the quasar spectra. This strategy is conceptually identical to the 
one discussed above for the Lyman-$\alpha$ lines of $HI$. However, in 
practice things are different. For, while positive detection would provide 
unambiguous evidence, sizable dust effects in the clouds would totally upset 
this method, and so no conclusion can be drawn from lack of detection. 

As a matter of fact, $H_2$ has already been detected in this way (Folz et al. 
1988; Ge \& Bechtold 1997), but these observations suffer from a severe 
confusion problem. Nevertheless, the {\it FUSE} satellite~(Sembach et al. 
2002) should collect high-quality data on the $H_2$ absorption lines, and so 
a clarification of this issue is to be expected.

{\it Clusters of galaxies} -- As is well known, clusters of galaxies contain 
a large amount of hot X-ray emitting gas. Besides thermal Bremsstrahlung, 
the intracluster gas also produces heavy-element recombination lines. This 
fact implies that at least part of the gas must have been processed inside 
the galaxies and subsequently ejected. Actually, specific 
models of galactic chemical evolution correctly account for the observed 
metallicity of the intracluster gas, but its predicted total amount 
invariably turns out to be about one order of magnitude {\it lower} than 
observed (Matteucci \& Vettotani 1988; Ciotti et al. 1991; Metzler \& Evrard 
1994). So, it appears that roughly $90 \, \%$ of the intracluster gas {\it 
cannot} be galactic in origin. 

To the extent to which the present dark matter 
scenario describes the cluster galaxies, such a conclusion is not necessarily 
true. In fact, what the above galactic evolution models implicitly assume is 
that {\it only luminous baryons} are present. However, a nontrivial amount of 
dark gaseous baryons lurking in galactic halos provides an additional supply, 
which can be progressively transferred to the intracluster medium via 
ram-pressure stripping and galaxy-galaxy interaction~\footnote{The relevance 
of galactic outflows for clusters of galaxies has been emphasized by Binney, 
Gerhard and Silk (2001).}. Moreover, an order-of-magnitude estimate shows 
that a large fraction of the intracluster gas can be accounted for in this 
way. Accordingly -- following a previous analysis~\footnote{This analysis was 
based on the preliminary evidence -- then turned out to be wrong -- that 
MACHOs make up $\sim 50 \%$ of the Milky Way halo.} by David~(1997) -- one 
can conclude that the fractional baryonic content of individual galaxies, 
groups and rich clusters is nearly a {\it constant} (independent of scale).

\section{Conclusions}

After some introductory remarks about the observational evidence for dark 
matter, we have shown that the need for baryonic dark matter is today even 
more solid than it was in the past. What is still missing, however, is some 
clear-cut observational evidence about the specific form taken by the dark 
baryons. 

Nevertheless, recent developments in computer simulations of galaxy formation 
as well as in the understanding of low-redshift Lyman-$\alpha$ clouds have 
provided valuable hints. When they are combined with metallicity and 
background light constraints, the list of baryonic dark matter candidates in 
galaxies gets dramatically shortened. 

Dark clusters of  brown dwarfs and cold $H_2$ clouds in galactic halos 
certainly look as a natural possibility. We have discussed how they are 
expected to form, the properties they presumably should have and the 
observations in which they are likely to show up. 

A good deal of theoretical and observational work has still to be done before 
claiming that the present dark matter scenario is correct. But -- we believe 
-- the stakes are worth the effort.

\section*{Acknowledgments}

It is a pleasure to thank M. Arnaboldi, G. F. Bignami, P. Caraveo, F. De 
Paolis, O. Gerhard, P. Gondolo, G. Ingrosso, Ph. Jetzer, J. Primack and E. 
Waxman for discussions, criticism and advice about the matter presented in 
this paper.

\end{document}